\documentclass[10pt,letterpaper]{article}
\usepackage[top=0.85in,left=2.75in,footskip=0.75in]{geometry}

\usepackage{amsmath,amssymb}

\usepackage{changepage}

\usepackage[utf8x]{inputenc}

\usepackage{textcomp,marvosym}

\usepackage{cite}

\usepackage{nameref,hyperref}

\usepackage[right]{lineno}

\usepackage{microtype}
\DisableLigatures[f]{encoding = *, family = * }

\usepackage[table]{xcolor}

\usepackage{array}

\newcolumntype{+}{!{\vrule width 2pt}}

\newlength\savedwidth

\raggedright
\usepackage[aboveskip=1pt,labelfont=bf,labelsep=period,justification=raggedright,singlelinecheck=off]{caption}

\makeatletter
\renewcommand{\@biblabel}[1]{\quad#1.}
\makeatother

\usepackage{lastpage,fancyhdr,graphicx}
\usepackage{epstopdf}
\fancyheadoffset[L]{2.25in}
\fancyfootoffset[L]{2.25in}
\newcommand{\new}[1]{{\color{black}#1}}
\newcommand{\newr}[1]{{\color{black}#1}}

\begin{document}
	\vspace*{0.2in}
	
	\begin{flushleft}
		{\Large
			\textbf\newline{Phenotypic-dependent variability and the emergence of tolerance in bacterial populations} 
		}
		\newline
		\\
		Jos\'e Camacho Mateu\textsuperscript{1, \new{\P}},
		Matteo Sireci\textsuperscript{2, \new{\P}},
		Miguel A. Mu\~noz\textsuperscript{2, *},
		\\
		\bigskip
		
		\textbf{1} Departamento de Matemáticas, Universidad Carlos III de Madrid, Leganés, Spain
		\\
		\textbf{2} Departamento de Electromagnetismo
		y F{\'\i}sica de la Materia and \\ Instituto Carlos I de F{\'\i}sica
		Te\'orica y Computacional. Universidad de Granada. Granada, Spain
		\\
		\bigskip
		
		%
		%
		\new{\P} These authors contributed equally to this work.
		
		* mamunoz@onsager.ugr.es
		
	\end{flushleft}
	\section*{Abstract}
	Ecological and evolutionary dynamics have been historically regarded
	as unfolding at broadly separated timescales.  However, these two
	types of processes are nowadays well-documented to \newr{intersperse} much
	more tightly than traditionally assumed, especially in communities of
	microorganisms. Advancing the development of mathematical and
	computational approaches to shed novel light onto eco-evolutionary
	problems is a challenge of utmost relevance. With this motivation in
	mind, here we scrutinize recent experimental results showing
	evidence of rapid evolution of tolerance by lag in bacterial
	populations that are periodically exposed to antibiotic stress in
	laboratory conditions.  In particular, the distribution of
	single-cell lag times
	—i.e., the times that individual bacteria from the community remain
	in a dormant state to cope with stress— evolves its average value to
	approximately fit the antibiotic-exposure time.  Moreover, the
	distribution develops right-skewed heavy tails, revealing the presence
	of individuals with anomalously large lag times. Here, we develop a
	parsimonious individual-based model mimicking the actual demographic
	processes of the experimental setup. Individuals are characterized by
	a single phenotypic trait: their intrinsic lag time, which is
	transmitted with variation to the progeny. The model —in a version in
	which the amplitude of phenotypic variations grows with the
	parent’s lag time— is able to reproduce quite well the key empirical
	observations. Furthermore, we develop a general mathematical framework
	allowing us to describe with good accuracy the properties of the
	stochastic model by means of a macroscopic equation, which generalizes
	the Crow-Kimura equation in population genetics. Even if the model
	does not account for all the biological mechanisms (e.g., genetic
	changes) in a detailed way —i.e., it is a phenomenological one— it
	sheds light onto the eco-evolutionary dynamics of the problem and can
	be helpful to design strategies to hinder the emergence of tolerance
	in bacterial communities. From a broader perspective, this work
	represents a benchmark for the mathematical framework designed to
	tackle much more general eco-evolutionary problems, thus paving the
	road to further research avenues.

	\section*{Author summary}
	
	Problems in which ecological and evolutionary changes occur at similar
	timescales and feedback into each other are ubiquitous and of outmost
	importance, especially in microbiology. A particularly relevant
	problem is that of the emergence of tolerance to antibiotics by lag,
	that has been recently shown to emerge very fast in bacterial
	(\newr{E. coli}) populations under controlled laboratory conditions. Here, we
	present a computational \newr{individual-based model, allowing
		us to reproduce empirical observations} and, also, introduce a very
	general analytical framework to rationalize such results. We
	believe that our combined computational and analytical approach may
	inform the development of well-informed strategies to mitigate the
	emergence of bacterial tolerance and resistance to antibiotics and,
	more generally, can help shedding light onto more general
	eco-evolutionary problems.

	\section*{Introduction}
	
	The extraordinary ability of species to adapt and survive in
	unpredictably-changing and unfavorable environments is certainly one
	of the most astonishing features among the many wonders of the
	phenomenon that we call life. Such adaptations can occur at extremely
	fast temporal scales thus interspersing ecological and evolutionary
	processes \cite{Hendry,Schoener,eco-evo}. A widely spread surviving
	strategy is \emph{latency} or \emph{dormancy}, i.e., the possibility
	for organisms to enter a period of reduced metabolic activity and
	non-replication adopted during adverse environmental conditions
	\cite{Lennon2011,Lennon2017,Fraser,
		guppy1999metabolic,sturm2015phenotypic}. Examples of dormancy can be
	found across kingdoms, with examples ranging from microorganisms such
	as viruses, bacteria or fungi
	\cite{barton2007herpesvirus,bertrand2019lag,van2015bacterial,normalcy}
	to plants \cite{Childs,Venable1} and animals
	\cite{guppy1999metabolic}.  During the latency period the organism is
	said to be in a \textit{latent or dormant state} and the time it takes
	to wake up is referred to as ``lag time'' or simply ``lag''. Entering
	and exiting a dormant state are not cost-free processes, since
	individuals may require of a specific metabolic machinery for
	performing such transitions and/or the development of
	specifically-devised ``resting structures''
	\cite{rechinger2000early,larsen2006differential,Lennon2011,Lennon2017,van2007microbial}.
	The exit from the dormant state can occur either as a response to
	environmental signals or cues
	\cite{Lennon2011,Lennon2017,Paredes2011,bertrand2019lag} or,
	alternatively, in a stochastic way
	\cite{Wright,Epstein2009,Buerger2012,Kussell-Leibler,Leibler-Kussell}. As
	a matter of fact, the duration of the lag intervals often varies
	widely between conspecific individuals and even between genetically
	identical organisms exposed to the very same environmental conditions
	\cite{Fridman,bertrand2019lag,Levin2017,Wright}. Such a variability is
	retained as an example of \emph{phenotypic diversification} or
	\textit{bet-hedging strategy} \cite{hedging,Kuipers} that confers a
	crucial competitive advantage in unpredictable and rapidly changing
	environments, thus compensating the above-mentioned individual costs
	and providing important benefits to the community as a whole
	\cite{Lennon2011,Lennon2017,bertrand2019lag,Villa2,xu2017phenotypic,Venable1}.

	Although, as already stated, latency is a widespread phenomenon,
	bacterial communities constitute the most suitable playground for
	quantitative analysis of latency owing to their diversity, fast life
	cycle, and the well-controlled conditions in which they can grow and
	proliferate in the laboratory \cite{kussell, Lenski,
		Lenski2,Elena}. Actually, latency was first described by Müller back
	in $1895$ as an explanation for the observed irregularities in the
	growth rate of bacterial cultures in his laboratory
	\cite{muller1895ueber}.  In recent years it has been realized that
	bacterial latency is a more complex and rich phenomenon than
	previously thought. Indeed, paraphrasing a recent review on the
	subject, the lag phase is ``dynamic, organized, adaptive, and
	evolvable'' \cite{bertrand2019lag}.
	
	Bacterial latency is at the root of \emph{tolerance} to antibiotics
	as, rather often, bactericidal antibiotics act during the reproduction
	stage and thus, by entering a dormant state, bacteria become
	transiently insensitive to antibiotics. Let us
	recall that bacterial \emph{tolerance} is not to be confused with
	bacterial \emph{resistance} \cite{Balaban2016}.  While
	\emph{resistance} refers to the ability of organisms to grow within a medium with antibiotics, provided these are not in high
	concentrations, \emph{tolerance} is the ability to transiently
	overcome antibiotics, even at very high concentrations, provided the
	exposition time is not too large \cite{kohanski2010antibiotics,
		Fridman, Balaban2016}. The strengths of these two
	complementary surviving strategies are quantified, respectively,  in terms of 
	quantities: {\bf(i)} the \textit{minimum inhibitory concentration}
	(MIC) of drug that must be supplied to stop the population growth ---a
	quantity that is significantly increased in resistant strains
	\cite{Balaban2016,Li2016importance,Levin2017, Fridman}--- and
	{\bf(ii)} the \textit{minimum duration to kill $99\%$ of the cells}
	$MDK_{99}$, which is increased in tolerant strains \cite{Balaban2017}.
	
	While the importance of bacterial resistance has long been recognized,
	studies underlining the crucial role played by tolerance are less
	frequent and more recent \cite{Balaban2016, Li2016importance,
		Levin2017, Fridman}.  An important caveat is that, while resistance
	is specific to one or a few antibiotics, tolerance is generically
	effective for a large diversity of them, leading to survival even
	under intensive multidrug treatment \cite{Balaban2016,Levin2017,
		Fridman}. Moreover, there exists firm evidence that tolerance is the
	first response to antibiotic stress \cite{Li2016importance},
	facilitating the later appearance of resistance
	\cite{Levin2017}. Therefore, understanding the emergence of tolerance
	is crucial for the development of more effective therapies aimed at
	dealing with recalcitrant infections and possibly preventing
	them. Aimed at shedding light on these issues, here we present an
	eco-evolutionary approach to analyze the emergence of tolerance by lag
	in bacterial communities under controlled laboratory experiments. In
	particular, we scrutinize the conditions under which modified lag-time
	distributions evolve as a response to stressful environments and
	investigate the origin of the experimentally-observed broad heavy
	tails in lag-time distributions (see below).
	
	Beside this specific focus, the present work has a broader breath. The
	example of rapid evolution of lag-time distributions is used as a test
	to prove a theoretical framework that we are presently developing.
	Our framework is similar in spirit to existing approaches such as the
	theory of \emph{``adaptive dynamics''} and related models in
	population genetics \cite{Geritz1,Geritz2, beginners,Dieckmann-Metz},
	but aims at reconciling and generalizing them.
	
	As a historical sidenote, let us recall that \emph{adaptive dynamics}
	(AD) was born as a generalization of evolutionary game theory
	\cite{Hofbauer2003} to allow for a set of strategies that is
	continuously varying and, upon which selection acts.  AD led to the
	satisfactory explanation of intriguing phenomena such as evolutionary
	branching \cite{Geritz1,Geritz2, Dieckmann+Law,DD3}, speciation
	\cite{DD1,DD2,Speciation}, diversification
	\cite{Doebeli-diversification,Doebeli-Coli}, the emergence of altruism
	and cooperation \cite{Doebeli-altruism, Doebeli-coop}, and the
	evolution of dispersal \cite{Doebeli-dispersal}. Importantly, its
	foundations are also mathematically well-established
	\cite{Champagnat}.  However, in spite of its very successful history,
	AD in its standard formulation has some limitations that make it not
	directly applicable to complex situations such as the one we aim at
	describing here:
	
	(i) First of all, in its standard formulation,
	populations are considered as monomorphic, i.e. point-like in
	phenotypic space; thus it does not allow for phenotypically-structured
	populations (see however \cite{Doebeli-distr, Doebeli-Rubin}).
	
	(ii) The
	``macroscopic equations'' of AD for the populations are not easily
	connected to microscopic birth-death processes in individual-based
	models \cite{Doebeli-mech}.
	
	(iii) Variations are assumed to be small, typically
	Gaussian-distributed and independent of the parent's phenotypic state.
	
	(iv) Variations are considered to be rare: ``after every
	mutational event, the ecological dynamics has time to equilibrate and
	reach a new ecological attractor'' \cite{Tunnels1}. In other words, a
	separation is assumed between ecological and evolutionary timescales,
	while in microbial communities, such processes may occur in 
	concomitance. Such a convergence of characteristic timescales is the
	hallmark of eco-evolutionary dynamics \cite{Hendry, Schoener, eco-evo}
	and is at the basis of fascinating phenomena such as eco-evolutionary
	tunneling \cite{Tunnels1, Tunnels2} and other rapid evolutionary
	phenomena \cite{Antibiotic_wild, cordero, Lenski2, Sanchez,
		Rapid-Fisher} which are difficult to account for in the standard
	formulation of adaptive dynamics.
	
	In what follows, we employ a theoretical framework in the spirit
	of statistical mechanics that aims to fill the gap between theory,
	phenomenological models and, most importantly, experiments.  Thus, our
	approach ---which is similar in spirit to previous work on bacterial
	quorum-sensing by E. Frey and collaborators
	\cite{Frey2017,Frey2018}--- implements a number of extensions with
	respect to standard AD \cite{geritz1997dynamics}, as it makes explicit
	the connection between individuals (microscale), and community
	dynamics (macroscale), introduces a general variation kernel, allows
	for large and phenotypic-state-dependent variations, etc.  These
	extensions allow us to study phenotypic diversity within a
	well-characterized eco-evolutionary framework. A full account of this
	general theoretical framework will be presented elsewhere
	\cite{Sireci}.  Let us finally emphasize that many of the above
	questions and extensions have been already tackled in the mathematical
	literature, at a formal level \cite{Champagnat, Buerger}. Yet, to the
	best of our knowledge, these results have have been confined to
	rigorous analyses of toy models and have not fully percolated through
	the biological and physical literature.
	
	The paper is organized as follows: in the first section we discuss in
	detail the experimental setup and empirical findings object of our
	study; then, we introduce a stochastic individual-based model
	implementing phenotypic variability and inheritability to account for
	experimental results. We present an extensive set of both
	computational and analytical results for it, discussing in particular
	the conditions under which the mathematical results deviate from
	computational ones. Finally, we discuss the implications of our work
	both from a biological viewpoint and how it contributes to the
	understanding of the evolution of heterogeneous phenotypic
	distributions, as well as from a more general eco-evolutionary
	perspective.

	\subsection*{Empirical observations: rapid evolution of lag-time distributions} 
	For the sake of concreteness, we focus on recent experimental results
	on the rapid evolution of tolerance in populations of
	\textit{Escherichia coli} in laboratory batch cultures in Balaban's
	lab \cite{Fridman}.  In particular, a bacterial population is
	periodically exposed to antibiotics (\textit{amplicillin}) in very
	high concentrations (much larger than the \textit{MIC}) during a
	fixed-duration time interval $T_a$ (e.g., $T_a=3, 5,$ or $8$
	hours). After antibiotic exposure the system is washed and the
	surviving population is regrown in a fresh medium during a time
	interval $T$ (with $T=23 h -T_a$). The antibiotics/fresh-medium cycle
	is iterated at least $8$ or $10$ times. Results are averaged over
	$2$ experimental realizations for each $T_a$ and the  resulting
	maximal carrying capacity is about $10^9$ individuals (we refer to \cite{Fridman} for
	further biological and experimental details).
	
	Once the cycles are completed, Fridman et al. \cite{Fridman}
	isolated some individuals from the surviving community and by
	regrowing them in a fresh medium they found that the distribution
	$P(\tau)$ of lag times $\tau$ ---i.e. the time individual dormant
	cells take to start generating a new colony after innoculation into a
	fresh medium--- changes from its ancestral shape to a modified one,
	shifted towards larger $\tau$ values. More specifically, the mean
	value grew to a value that approximately matches the duration of the
	antibiotic-exposure time interval, $T_a$ (see 
	\cite{Fridman}). This modified lag-time distribution entails an
	increase in the survival probability under exposure to ampicillin
	but, also, to antibiotics of a different bactericidal class such as
	\textit{norfloxacin}, for the same time period. Furthermore,
	mutations were identified in diverse genes, some of them known to be
	related with regulatory circuits controlling the lag-time
	distribution, such as the toxin{\textendash}antitoxin one
	\cite{Toxin_Balaban}.  Subsequently, after many cycles, the population
	was also observed to develop resistance to ampicillin \cite{Fridman}.
	Thus, the conclusion is that non-specific tolerance ---stemming from
	lag--- emerges in a very rapid way as a first adaptive change/response
	to antibiotic stress. More in general, these results reveal that
	the adaptive process is so fast that ecological and evolutionary
	processes occur at comparable timescales
	\cite{Rapid-Fisher,Rapid2016,Bonachela1,Bonachela2,Bonachela3}.
	
	The experimentally-determined lag-time distributions reveal another
	intriguing aspect that ---to the best of our knowledge--- has not been
	extensively analyzed so far: their variance is also significantly
	increased as $T_a$ grows and, related to this, the resulting mean
	value of the distribution is always larger than its median
	\cite{Balaban_scan}. This is an indication that, as a matter of fact,
	the empirically-obtained lag-time distributions are skewed and exhibit
	heavy tails, including phenotypes with anomalously-large lag times
	---much larger than $T_a$---, especially for large $T_a$'s.  This
	observation is surprising as, under such controlled lab conditions,
	one could naively expect to find lag-time distributions sharply peaked
	around the optimal time value, $T_a$, since, ideally, the best
	possible strategy would be to wake-up right after antibiotics are
	removed and any further delay comes at the price of a reduction of the
	overall growth rate or fitness. Fridman \textit{ et al.}  proposed
	that the increase in the variance might suggest a past selection for a
	bet-hedging strategy in natural unpredictable environments; however,
	anomalously-large lag-time values were not present in the original
	wild-type population. The authors also suggested that there could be
	constraints at the molecular level imposing the mean and the variance
	of the lag-time distribution to increase concomitantly \cite{Fridman,
		Toxin_Balaban}, a possibility that inspired us and that we will
	carefully scrutinize from a theoretical and computational perspective
	in what follows.
	
	\section*{Model building }\label{ch:Model}
	\begin{figure}[t]
		\centering
				\includegraphics[width=\textwidth]{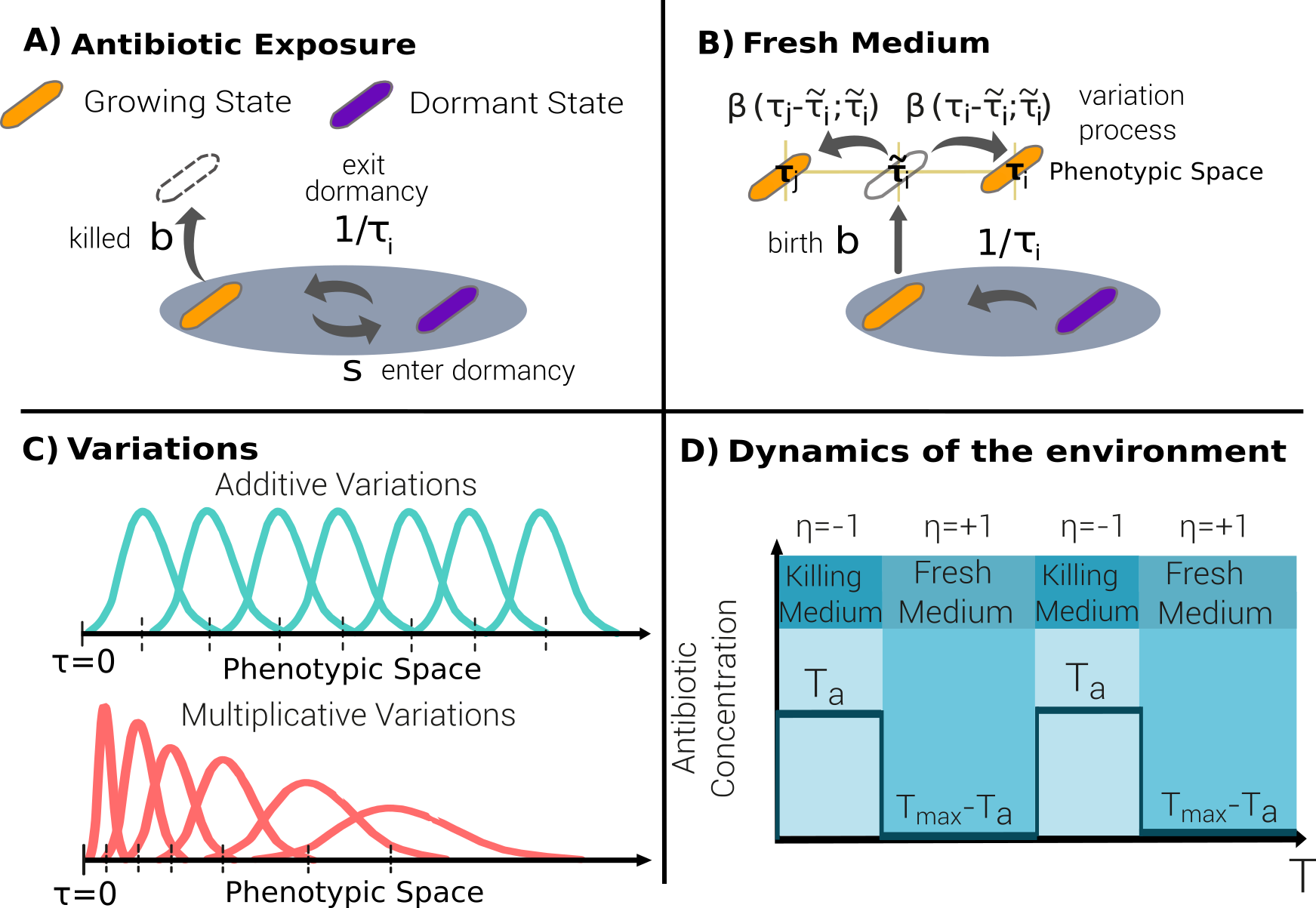}
		\caption{\textbf{Sketch of the main ingredients of the
				individual-based stochastic model}. Each individual bacterium ($i$)
			is characterized by its phenotypic state, lag time $\tau_i$ and
			experiences demographic processes.  \textbf{(A)} In the presence
			of antibiotics, bacteria can stochastically switch between the dormant and the
			growing state (at transition rates $s$ and $1/\tau_i$, respectively);
			growing individuals can also attempt reproduction (at a ``birth''
			rate $b$) and be immediately killed by the action of antibiotics
			(as bactericidal antibiotics usually act during duplication
			attempts).  \textbf{(B)} In the fresh medium, dormant bacteria
			can wake up at a rate $1/\tau_i$, that depends on their
			intrinsic (phenotypic) lag time; on the other hand, growing
			cells can reproduce asexually by duplication; the resulting
			offspring inherit the characteristic time scale with some
			variation, as specified by a function $\beta$. \textbf{(C)} Two
			possible types of variation functions $\beta$: in the additive
			case (top), the standard deviation is constant, i.e. independent
			of the initial state $\tau_i$, while in the multiplicative case
			(bottom) the standard deviation is assumed to grow linearly with
			the parent's lag time $\tau_i$. \textbf{(D)} Sketch of the
			environmental variation, alternating periodically between
			antibiotic exposure (time $T_a$) and a fresh medium
			($T_{max}-T_a$). } \label{Fig 1}
	\end{figure}
	
	Aimed at shedding light onto these empirical findings, here we
	propose an individual-based stochastic model for phenotypic
	adaptation in which each single individual cell can be either in an
	``awake'' or in a ``dormant'' state \cite{Balaban,
		Kussell-Leibler,Leibler-Kussell,Baranyi2002,Baranyi2005} (see
	Fig \ref{Fig 1} for a sketch of the model).  Mimicking the
	experimental protocol of Fridman et al.--- a population of such
	individuals is exposed to alternating adverse and favorable
	conditions with durations $T_a$ and $23h - T_a$, respectively (a
	function $\eta(t)$ labels the environmental state at any given
	time $t$: $\eta(t)=-1$ in the presence of antibiotics and
	$\eta(t)=+1$ in the fresh medium).
	
	The model assumes that each awake cell is able to sense the
	environment and respond to it by regulating its state: they can
	sense the presence of antibiotics and enter the dormant state at
	rate $s$, while such a machinery is
	assumed to be turned off during dormancy.  In S1 Text, Sec. S5, we also consider a
	generalization of the model in which awake individuals can also
	enter the dormant state as a response to other sources of stress
	such as starvation \cite{Wide}.  Indeed, the wake-up is
	assumed to occur as a result of a Markovian stochastic process; each
	individual bacteria $i$ is \emph{phenotypically} characterized by
	its intrinsic typical mean lag time $\tau_i$ meaning that, it wakes
	up stochastically at a constant transition rate
	$1/\tau_i$. Therefore, the time $t$ at which a dormant cell actually
	resumes growth is a random variable distributed as
	$P(t\vert \tau_i)=e^{-t/\tau_i}/\tau_i$, with mean value $\tau_i$
	\cite{vKampen,Gardiner}.  In the last section we discuss recent
	alternatives to Markovian processes, i.e. including some form of
	``memory'', which can give rise to non-exponential residence
	times, to describe this type of waking-up phenomena
	\cite{Norman-review,Mitarai}. 
	
	Awake individuals are exposed to stochastic demographic processes:
	they attempt asexual reproduction (i.e., duplication) at a constant
	birth rate $b$ and die spontaneously at rate $d$ (that we fix to $0$
	without loss of generality). Reproduction attempts are successful in
	the fresh medium while, in the presence of antibiotics, they just lead
	to the parent's death and its removal from the community.  Following
	this dynamics, the population can freely grow, until its size reaches
	a maximal carrying capacity $K$. Once this limit has been reached, the
	population enters a saturated regime, within which each new birth is
	immediately compensated by a random killing (much as in the Moran
	process \cite{moran_1958}).  
	
	Importantly, in parallel with the above demographic processes, the
	model implements an evolutionary/adaptive dynamics.  The phenotypic
	state $\tau_i$ of each successfully dividing individual is
	transmitted, with possible variation, to its progeny. In particular,
	the two offspring resulting from duplication have phenotypic states
	$\tau_i+\xi_1$ and $\tau_i+\xi_2$, respectively, where $\xi_1$ and
	$\xi_2$ are the phenotypic stochastic variations, sampled from some
	probability distribution, that we generically call
	$\beta(\xi; \tau_i)$ and that, in the more general case, can be
	state-dependent, i.e.  depend on $\tau_i$. More specifically, we
	implemented two different variants of the model, depending of the
	standard deviation of the probability distribution
	$\beta(\xi;\tau_i)$:
	\begin{itemize}
		\item The \emph{additive} model, with a standard deviation, $\alpha_A$, common to all phenotypes.
		
		\item  The \emph{multiplicative} model, with a state-dependent standard deviation, $\alpha_M \tau_i$,
		for individuals with intrinsic lag time $\tau_i$, where $\alpha_M$ is a constant (see Methods).
	\end{itemize}
	Observe that in the multiplicative case, the larger the parent's lag time
	the larger the possible amplitude of variations, in a sort of
	rich-get-richer or Matthew-effect mechanism, well-known in the theory
	of stochastic processes to generate heavy tails \cite{Sornette,
		Barabasi, Mitzenmacher, Newman,
		MN-Manrubia,MN1,MN2,MN-Munoz,MN3,Genovese}. As a motivation for this choice,
	let us mention that there is solid evidence
	that the genetic circuits involved in the regulation of the lag-time
	distribution (such as the toxin-antitoxin one), can indirectly produce
	this type of fluctuations at the phenotypic level
	\cite{Toxin_Balaban}. Furthermore, similar phenotypic-variation
	kernels have been argued to arise from non-linear effects in the way
	genotypic changes (mutations) are manifested into phenotypic
	variability (see e.g. \cite{van,Pascual}).

	\section*{Analytical (mean-field) theory}
	\label{ch:framework}
	
	Before delving into computational analyses of the model, let us
	present a mathematical framework allowing us to obtain
	theoretical insight.  Readers not particularly interested in
	analytical approaches can safely skip this section, and just be aware
	that it is possible to mathematically understand all the forthcoming
	computational results.
	
	The previous Markovian stochastic individual-based model is
	mathematically defined as a ``many-particle'' Master equation ruling
	the time evolution of the joint probability-distribution functions for
	the whole set of all ``particles'' ( i.e., cells). The resulting
	master equation can be simulated computationally by employing the
	Gillespie Algorithm (see below and S1 Text, Sec. S2A, for details)
	\cite{vKampen,Gardiner,Gillespie}. However, as it is often the case
	for such many-particle Master equations, it is hard to handle analytically in
	an exact way.  Thus, in order to gain quantitative understanding
	beyond purely computational analyses, here we develop an approximation
	---which becomes exact in the limit of infinitely large population
	sizes \cite{Frey2017,Frey2018}--- that allows us to derive a
	macroscopic (or ``mean-field'') description of the stochastic model in
	terms of the probability density of finding an individual at any given
	phenotypic state, $\tau$ (i.e. the ``one-particle'' probability
	density).  The mean-field approach that we employ in what follows is
	just a first example of a much more general framework that we will
	expose in detail elsewhere \cite{Sireci}.

	A first step toward the derivation of a \emph{macroscopic equation}
	relies on a marginalization of the many-particle
	probability-distribution function to obtain a one-particle probability
	density (see S1 Text, Sec. S2B). The resulting marginalized distribution function
	encapsulates the probability density $\phi(\tau,t)$ that a randomly
	sampled individual at time $t$ has lag time $\tau$. This probability
	---that needs to be normalized, so that
	$\int^{\infty}_{0} \phi(\tau, t)d\tau =1$--- can be decomposed in two
	contributions $\phi(\tau,t)=\phi_{G}(\tau,t)+\phi_{D}(\tau,t)$
	representing, respectively, the relative fraction of individuals in
	growing (G) and dormant (D) states. Observe that these two densities
	are not probability distributions and thus they are not normalized to
	unity separately.  In the limit of infinitely-large population sizes,
	the evolution of the probability density for individuals in the
	growing state, ${\phi}_G(\tau , t) $ is ruled by the following equation (details of the
	derivation can be found in SI, S2A):
	\begin{eqnarray}
	\partial_t \phi_G(\tau , t) &=&\frac{1+\eta(t)}{2}\left[-b\phi_G(\tau, t) +2b\int^{\infty}_{0} d\tilde{\tau}\beta(\tau-\tilde{\tau};\tilde{\tau})\phi_{G}(\tilde{\tau}, t) \right.\nonumber\\ 
	&-&\left. b \phi_G(\tau, t)\int^{\infty}_{0} d\tilde{\tau}\phi_{G}(\tilde{\tau},t) \right] \nonumber\\
	&-&\frac{1-\eta(t)}{2}\left[ b\Big( 1-\int^{\infty}_{0} d\tilde{\tau}\phi_{G}(\tilde{\tau},t)\Big)
	+s\right] \phi_G(\tau,t) +\frac{1}{\tau}\phi_D(\tau,t). \label{eq:growing-mf1}
	\end{eqnarray}
	Even if this equation might look cumbersome, its different terms have a rather intuitive interpretation:
	
	\begin{itemize}
		\item \emph{In the fresh medium} (terms proportional to $1+\eta(t)$):
		(i) the first term represents the negative probability flow stemming
		from growing individuals with generic phenotypic trait $\tau$ that
		reproduce (at rate $b$) and change to any other arbitrary phenotypic
		state; (ii) the second represents the positive contribution of
		reproducing individuals (at rate $b$) with any arbitrary trait
		$\tilde{\tau}$, for which one of the two resulting offspring jumps
		to $\tau$ (controlled by the function
		$\beta(\tau-\tilde{\tau};\tilde{\tau})$); (iii) the third
		\emph{selection term} stems from the normalization of the overall
		probability density: if the population size grows because any
		individual with arbitrary trait $\tilde{\tau}$ reproduces
		successfully (at rate $b$), then the relative probability to observe
		phenotype $\tau$ decreases to keep the overall probability-density
		conserved.
		
		\item \emph{ In the presence of antibiotics} (terms proportional to $1-\eta(t)$): 
		(i) the first term represents the rate at which growing individuals that attempt reproduction (at rate $b$) are killed by antibiotics; (ii) the second term is a selection term, fully analogous to the above-discussed one: when any arbitrary individual dies the overall probability density at $\tau$ increases;  (iii) the third term represents the outflow of individuals entering the dormant state at rate $s$. 
		
		\item \emph{In both environments} (no dependence on $\eta(t)$): the only
		term, proportional to the rate $1/\tau$, describes the probability
		inflow stemming from dormant individuals that become awake.
	\end{itemize}

	Similarly, the equation for the density of individuals in the dormant state is
	\begin{eqnarray}\label{eq:dormant-mf}
	\partial_t \phi_D(\tau, t)&=& 
	-\eta(t) b \phi_D(\tau,t) \int^{\infty}_{0}d\tilde{\tau}\phi_{G}(\tilde{\tau},t) - \frac{1}{\tau}\phi_D(\tau,t)
	+\frac{1-\eta(t)}{2}s\phi_G(\tau,t)
	\end{eqnarray}
	where the first (selection) term stems from the overall probability
	conservation when the population either  grows or shrinks (negative or
	positive signs, respectively), and the remaining two terms have the opposite meaning (and signs) of their
	respective counterparts in Eq.(\ref{eq:growing-mf1}).
	
	In order to make further analytical progress, in the case in which
	variations are assumed to be small, it is possible to introduce a
	further (``diffusive'' or ``Kimura'') approximation as often done in
	population genetics as well as in adaptive or evolutionary
	mathematical approaches \cite{Kimura-diffusion}.  More specifically,
	one can perform a standard (Kramers-Moyal) expansion of the master
	equation by assuming that jumps in the phenotypic space are relatively
	small \cite{vKampen,Gardiner}, i.e. expanding the function $beta$ in
	Taylor series around $0$. After some simple algebra (see S1 Text, Sec. S3) one obtains a
	particularly simple expression for the overall probability
	distribution:
	
	\begin{eqnarray}\label{GCK}
	\partial_t \phi(\tau , t)&=&\eta(t)\left[f(\tau , t)-\bar{f}( t)\right]\phi(\tau , t)\nonumber\\
	&-&(\eta(t)+1) \left[ \partial_{\tau}\theta(\tau)f(\tau,t)\phi(\tau,t)
	-\frac{1}{2}\partial_\tau^2\sigma^2(\tau)f(\tau,t) \phi(\tau,t) \right]
	\end{eqnarray}
	where the \emph{`` effective fitness function''}
	$f(\tau,t)\equiv b \phi_G (\tau , t)/\phi (\tau , t)$ and its
	population average
	$\bar{f}(t)=\int^{\infty}_{0}d\tau f(\tau,t)\phi(\tau,t)$ have been
	introduced, and where $\theta(\tau) $ and $\sigma^2(\tau) $ are the
	first and second cumulants of the variation function $\beta$ (in first
	approximation we can assume $\theta(\tau)=0$, while
	$\sigma^2(\tau)=\alpha_A^2$ for the additive case and
	$\sigma^2(\tau)=\alpha_M^2 \tau^2$ for the multiplicative
	case). Observe that, remarkably, this last equation is a
	generalization of the celebrated continuous-time \emph{Crow-Kimura
		equation} of population genetics \cite{crow1970introduction}, also
	called \emph{selection-mutation equation}
	\cite{Hofbauer,Page+Nowak}. In particular, notice that the dynamics of
	the probability density is exposed to the combined action of the
	process of selection (first term in Eq.(\ref{GCK}), which is nothing
	but the \emph{replicator equation} \cite{Hofbauer2003,nowak2006evolutionary})
	and mutation, as specified by the drifts in the second line. This type
	of equations, combining replicator dynamics with Fokker-Planck type of
	terms ---even if with a slightly different interpretation--- have been
	also studied by Sato $\&$ Kaneko and Mora $\&$ Walzak
	\cite{Kaneko2006,Kaneko2007,Mora}.  The main ---and crucial---
	differences between Eq.(\ref{GCK}) and the standard Crow-Kimura
	equation are:
	\begin{itemize}
		
		\item The fitness function appears in the mutation terms ---whereas in the standard Crow-Kimura equation the diffusion term would  read $\partial^2_\tau\phi(\tau)$--- thus correlating reproduction rates and mutation amplitudes. Observe that here variations are always associated with reproduction events, as typically in bacteria and viruses, in such a way that a higher fitness rate implies a higher mutation rate.
		
		\item There is a  general dependence on the cumulants of the variation kernel that, in general, can be trait-dependent and asymmetric. 
		
	\end{itemize}
	These generalizations are essential ingredients to capture the essence
	of our Markovian model as we will see and, to the best of our
	knowledge, have not been carefully analyzed in the past.
	From here on, we refer to Eq.(\ref{GCK}) as the \emph{generalized
		Crow-Kimura (GCK) equation}.
	\section*{Results} \label{ch:results} 
	
	\begin{figure}
		\centering
				\includegraphics[width=0.7\textwidth]{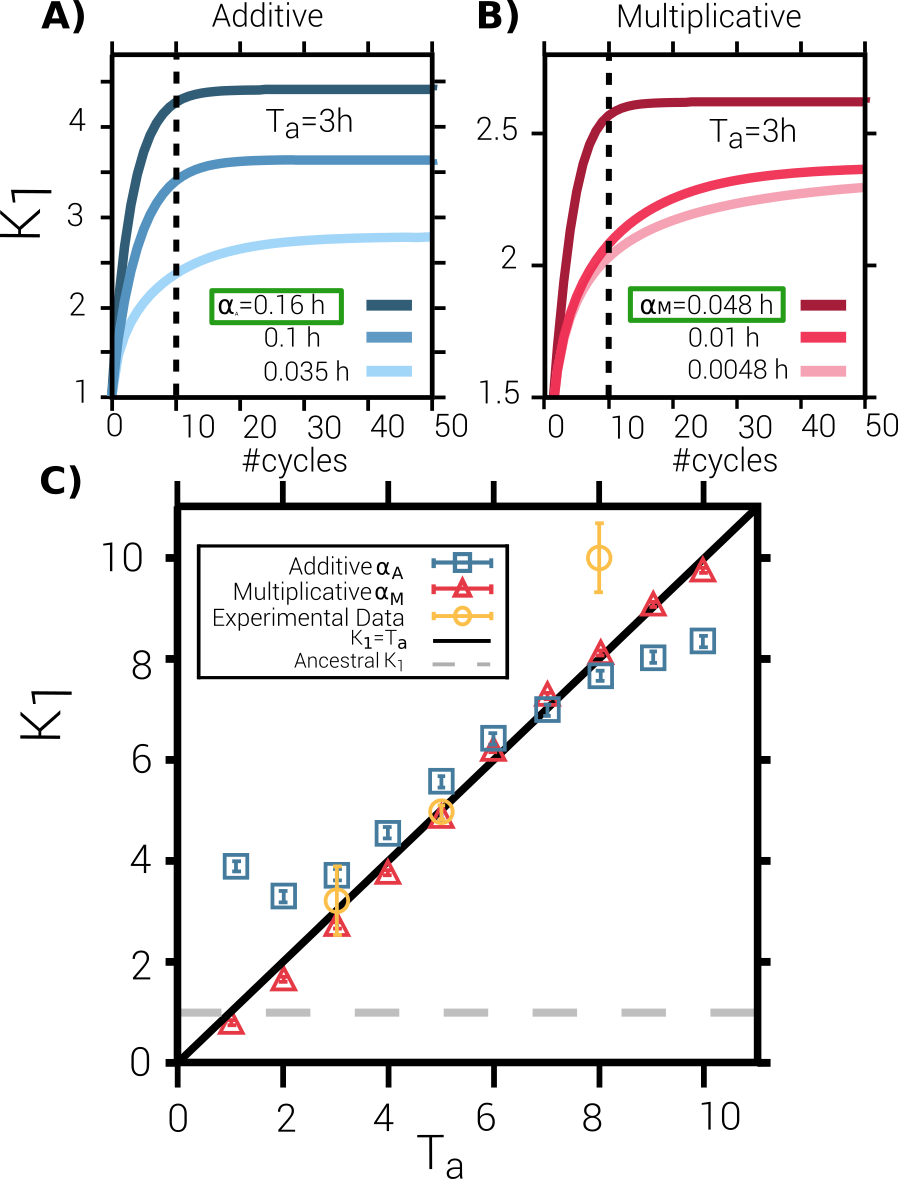}
		\caption{\textbf{Tuning the only free parameter to match empirical
				results}.\textbf{(A-B)} Mean of the lag-time distribution
			$K_1$, measured right at the end of each antibiotic cycle, as a
			function of the number of cycles. Computational results are
			shown for antibiotic duration $T_a=3 h$ for both the additive
			(A) and the multiplicative (B) versions of the model.  The
			different curves (color coded) correspond to different mutation
			amplitudes $\alpha_A$ (in A) and $\alpha_M$ (in B),
			respectively. The dashed vertical lines indicate the $10th$
			cycle, when experiments stop. Remarkably, the mean lag time
			strongly depends on the mutational amplitude, both in the
			transient regime and in the asymptotic state. We implement an
			algorithmic search to tune the only free parameter (either
			$\alpha_A$ or $\alpha_M$) to best fit the experimental mean lag
			times for all values of $T_a$ together and, in particular, their
			experimentally-reported linear dependence on the antibiotic
			exposure time $T_a$ (see Methods). \textbf{(C)} Mean of the
			lag-time distribution as a function of $T_a$ for the model
			(squares for additive and triangles for multiplicative versions
			of the model) tuned to reproduce experimental values (yellow
			symbols).  While empirical data are available for $T_a=3, 5$ and
			$8 h$, the model can be analyzed for generic values of $T_a$.
			The solid line indicates the linear dependence between the mean
			and lag-time distribution, $K_1=T_a$, while the horizontal
			dashed line represents the mean lag time of the ancestral
			population. Parameter values: $K=10^5$, $\alpha_{A}=0.16 h$,
			$\alpha_{M}=0.048 $, $b=2.4 h^{-1}$,
			$d=3.6\cdot10^{-5} h^{-1} $, $s=0.12\ h^{-1}$,
			$T_{fresh}=23 h-T_a$, $10$ cycles (see Methods).}
		\label{Fig 2}
	\end{figure}
	\begin{figure}
		\centering
				\includegraphics[width=\textwidth]{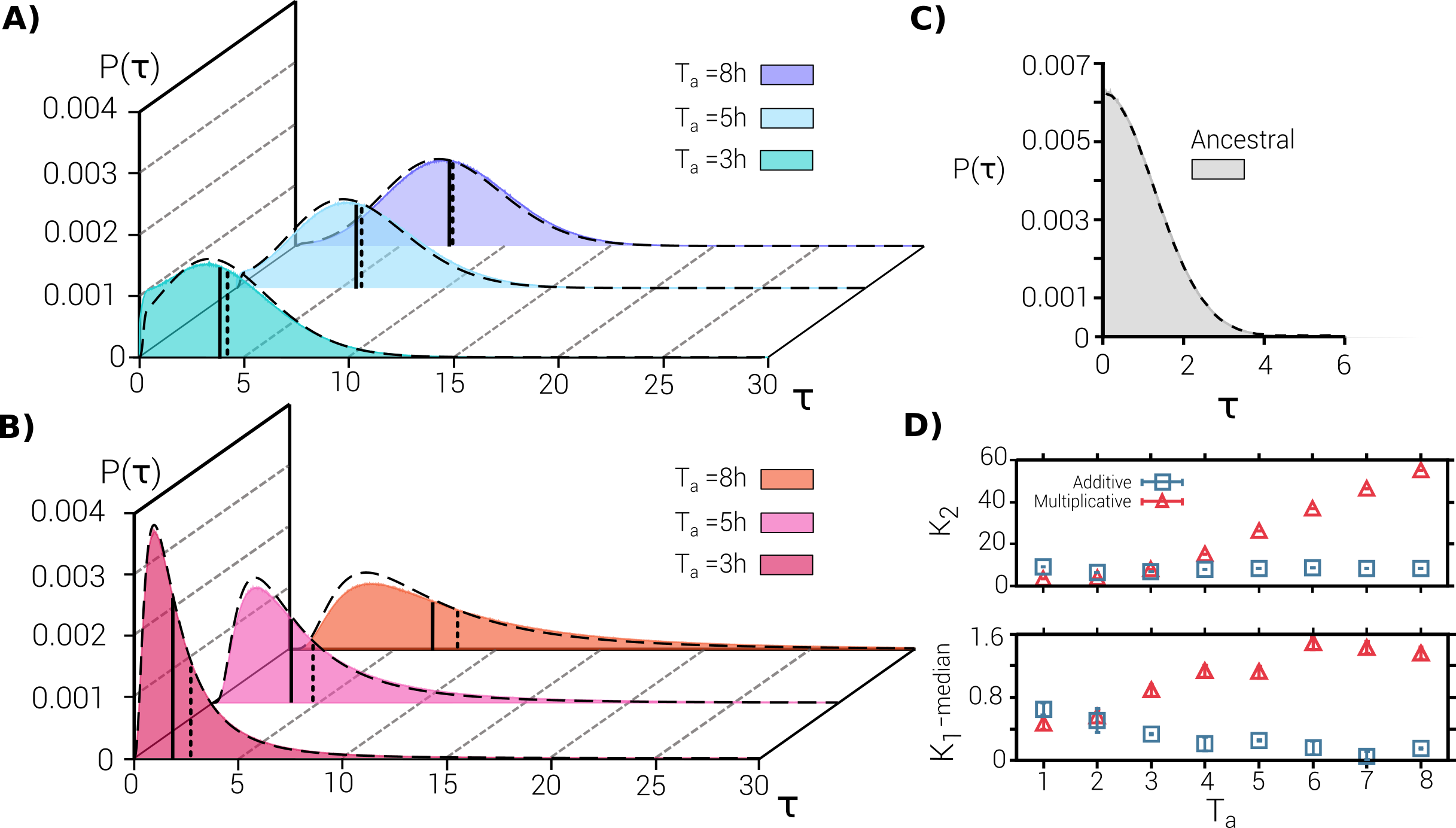}
		\caption{\textbf{Lag-time probability distributions: theory and
				simulations.} \textbf{(A-B)} Lag-time distribution after $10$
			cycles as obtained in the simulation of the individual-based
			model in both the additive (A) and the multiplicative (B) case,
			for different antibiotic-exposure periods, $T_a=3, 5$ and $8 h$
			(\newr{marked with different colours}).  Solid and dotted
			vertical lines indicate, respectively, the median and mean of
			the corresponding distributions (a large separation between
			these two indicators reflects asymmetries in the distribution
			such as the emergence of a heavy tail to the right).  Dashed
			lines represent results from the numerical integration of the
			GCK equation, Eq.\ref{GCK}, using the same parameters and
			external conditions. Observe that the multiplicative model
			generates much larger tails, reproducing the experimental
			phenomenology better than the additive one. \textbf{(C)} Initial
			lag-time distribution mimicking the experimentally observed one
			for the ancestral population.  \textbf{(D)} Variance, $K_2$, and
			difference between mean and median, $K_1-median$, of the
			lag-time distribution as a function of $T_a$ in the additive
			(blue squared symbols) and multiplicative (red triangular
			symbols) versions of the model.  $K_2$ grows with the antibiotic
			exposure time in the multiplicative case, while in the additive
			case it remains nearly constant.  The difference between the
			mean and the median is very small in the additive case, while it
			increases with $T_a$ almost monotonically in the multiplicative
			one. In summary, the multiplicative model generates a
			distribution with a variance that grows with the mean, as well
			as heavy tails, reproducing well the key experimental
			findings. Parameter values are as in Fig
			\ref{Fig 2}.}  \label{Fig 3}
	\end{figure}
	
	In order to scrutinize whether the proposed adaptive stochastic model
	can account for the key empirical findings of Fridman \emph{et al.}
	\cite{Fridman}, we perform both (i) extensive computational
	simulations and (ii) numerical studies of the mean-field macroscopic equation,
	Eq.(\ref{GCK}).
	
	\begin{itemize}
		\item Computational simulations rely on the Gillespie algorithm \cite{Gillespie}, which allows us to simulate exactly the master equation defining the stochastic model. In all cases, we consider at least $10^3$ independent realizations to derive statistically-robust results. Without loss of generality and owing to computational costs, the maximal population size or carrying capacity is fixed to $K=10^5$.
		
		\item On the other hand, for analytical approaches, in spite of the
		relatively simple form of Eq.(\ref{GCK}) owing to its non-linear
		nature and to the time-variability of environmental conditions
		$\eta(t)$, it is not possible to solve it analytically in a closed
		way and, thus, it becomes mandatory to resort to
		numerical-integration schemes. In particular, from this equation
		---or, more precisely, from integration of its two additive
		components: Eq.(1) and Eq.(2)--- one can derive the time-dependent
		as well as the asymptotic lag-time distributions 
		and, from them, monitor the leading moments or cumulants as a
		function of time.
	\end{itemize}
	Further details of both computational simulations and numerical
	integration of the macroscopic equation can be found in the Methods
	section as well as in the S1B. In what follows we present together
	both types of analyses, underlining where the mean-field approach
	works well and where its predictions deviate from direct
	simulations.

	\subsection*{Transient dynamics: determining variational amplitudes}
	
	Parameter values in the model are fixed to agree as much as possible
	with the empirical ones measured by Fridman et al. \cite{Fridman} (see
	Methods). In particular, we used (i) the same set of
	environmental-period durations $T_a$ and $T_{max}-T_a$ as in the
	antibiotics/fresh-medium cycle, (ii) the experimentally measured
	reproduction rate in the fresh medium, (iii) the empirical
	``falling-asleep'' rate $s$, as well as (iv) the same number of
	antibiotic cycles (ten) as in the experimental setup.  Initially all
	individuals are assumed to have small intrinsic lag-time values of
	$\tau$. In particular, we consider a truncated normal distribution
	with mean value and variance as in the actual ancestral population in
	the experiments ($\langle \tau \rangle^{exp.} =\ 1.0\ \pm0.2\ h$).
	
	Employing this set of experimentally-constrained parameter values and
	initial conditions, we ran stochastic simulations in which the whole
	population expanded and then shrank following the periodically
	alternating environments.  Along this dynamical cyclic process the
	distribution of $\tau$ values across the population varies in time; in
	particular, we monitored the histogram of $\tau$ values and obtained
	the corresponding probability distributions right at the end of each
	antibiotic cycle, just before regrowth, as in the experiments.
	
	Fig \ref{Fig 2}A and \ref{Fig 2}B show the evolution of
	the mean (i.e. the first cumulant, $K_1$) across cycles, while
	Fig \ref{Fig 3}A and \ref{Fig 3}B illustrate the full
	distribution and higher-order cumulants after $10$ cycles.
	Observe that the value of $K_1$ after $10$ cycles depends on
	the choice made for the only remaining free parameter, i.e.
	the variation-amplitude parameter $\alpha_A$ or $\alpha_M$,
	for the additive or multiplicative versions of the model,
	respectively.  In order to tune either of these parameters, we
	imposed that $K_1(10)$ reproduces in the closest way the
	experimentally determined values, as measured right before the
	$10th$ regrowth cycle. This tuning procedure leads to
	$\alpha_A=0.16(1) h$ and $\alpha_M=0.048(1)$ for the additive
	and multiplicative cases, respectively (parentheses indicate
	uncertainty in the last digit); these are the values that best
	reproduce the empirical findings in the sense of least-square
	deviation from the available empirical data for different
	$T_a$'s (see Fig \ref{Fig 2}C).

	Let us remark that both variants of the model are able to reproduce
	the key experimental feature of generating mean lag times close to
	$T_a$ (observe, however, that there is always a small deviation in the
	case $T_a =8 ~h$, for which even experimentally,
	$K_1 \approx 10 ~h > T_a$). Nevertheless, as illustrated in Fig
	\ref{Fig 3}A and \ref{Fig 3}B there are significant differences between the
	two variants. In particular, the additive model fails to reproduce the
	following empirical observations:
	\begin{itemize}
		\item $T_a$-dependent variances, 
		
		\item large differences between median and mean values, and 
		
		\item  strongly skewed distributions with large tails.
	\end{itemize}
	
	For instance, in the experiments, for $T_a=8 ~h$, the difference
	between the mean and the median is $1.1(1) ~h$ while in the additive
	model is $0.15 ~h$, i.e. about one order of magnitude
	smaller. Furthermore, in the experiments, lag times of up to $30 ~h$
	are observed, while in the additive model values above $\approx 15$
	are exponentially cancelled; i.e. they have an extremely low
	(negligible) probability to be observed.  This is also illustrated in
	Fig \ref{Fig 3}D where the second and third cumulants
	(variance and skewness) of the distribution after $10$ cycles are
	plotted as a function of $T_a$. Observe that both cumulants remain
	almost constant, revealing the absence of heavy tails for large values
	of $T_a$.
	
	On the other hand, the multiplicative model is able to reproduce not
	only the experimental values of the mean but also ---with no
	additional parameter nor fine tuning--- (i) the existence of large
	lag-time variances that increase with $T_a$, (ii) the above-mentioned
	large differences between the mean and the median ($1.3(1)$ in this
	case), as well as (iii) heavily skewed lag-time distributions that
	strongly resemble the empirically measured ones (see \new{Fig 2} in
	Fridman \emph{et al.} \cite{Fridman}). In particular, lag times of the
	order of $30 ~h$ have a non-negligible probability to be observed for
	$T_a=8 ~h$ within the multiplicative version of the model, after $10$
	cycles. The resulting probability and the corresponding cumulants (see
	Fig \ref{Fig 3}D depend strongly on $T_a$.
	
	Importantly, the previous results are quite robust against
	changes in the model.  In particular, if growing cells are allowed
	to switch to dormancy in response of starvation, the mean lag time
	increases, as expected, but the qualitative shape of the lag-time
	distribution remains unchanged (see S6 ). Hence, just by modifying
	accordingly the parameter $\alpha$ allows one to recover the same
	conclusions.
	
	As a word of caution let us emphasize that the distributions in
	Fig \ref{Fig 3} are not obtained exactly in the same
	way as the experimental ones. The first are distributions of 
	characteristic times $\tau$ (inverse of intrinsic transition rates)
	while the second are the actual lag
	times $t$ measured after regrown in a fresh medium. Actually,
	the characteristic time $\tau$, in our model, is just a
	proxy for the actual time that it takes for the colony formed by
	such an individual to be observable or detectable in actual experimental
	setups. Below we discuss this issue more extensively as well as the
	possible limitations it implies and extensions of the modelling
	approach to circumvent them. 
	
	Let us also underline that Fig \ref{Fig 3} reports not only
	the results of direct simulations but also the theoretical predictions
	(dashed lines) derived from numerical integration of the macroscopic equations
	for the two different cases. The agreement with simulation results is
	remarkably accurate; the origin of the existing small discrepancies
	will be analyzed in detail in a forthcoming section.

	Thus, the main conclusion of these computational and theoretical
	analyses is that \emph{state-dependent (multiplicative) variability is
		needed in order to account for the empirically observed key features
		of the lag-time distributions emerging after a few
		antibiotic/fresh-medium cycles.} Once this variant of the model is
	chosen, a good agreement with experimental findings
	if obtained by fitting the only free parameter: the amplitude of
	variations.

	\begin{figure}
		\centering
		
		\includegraphics[width=\textwidth]{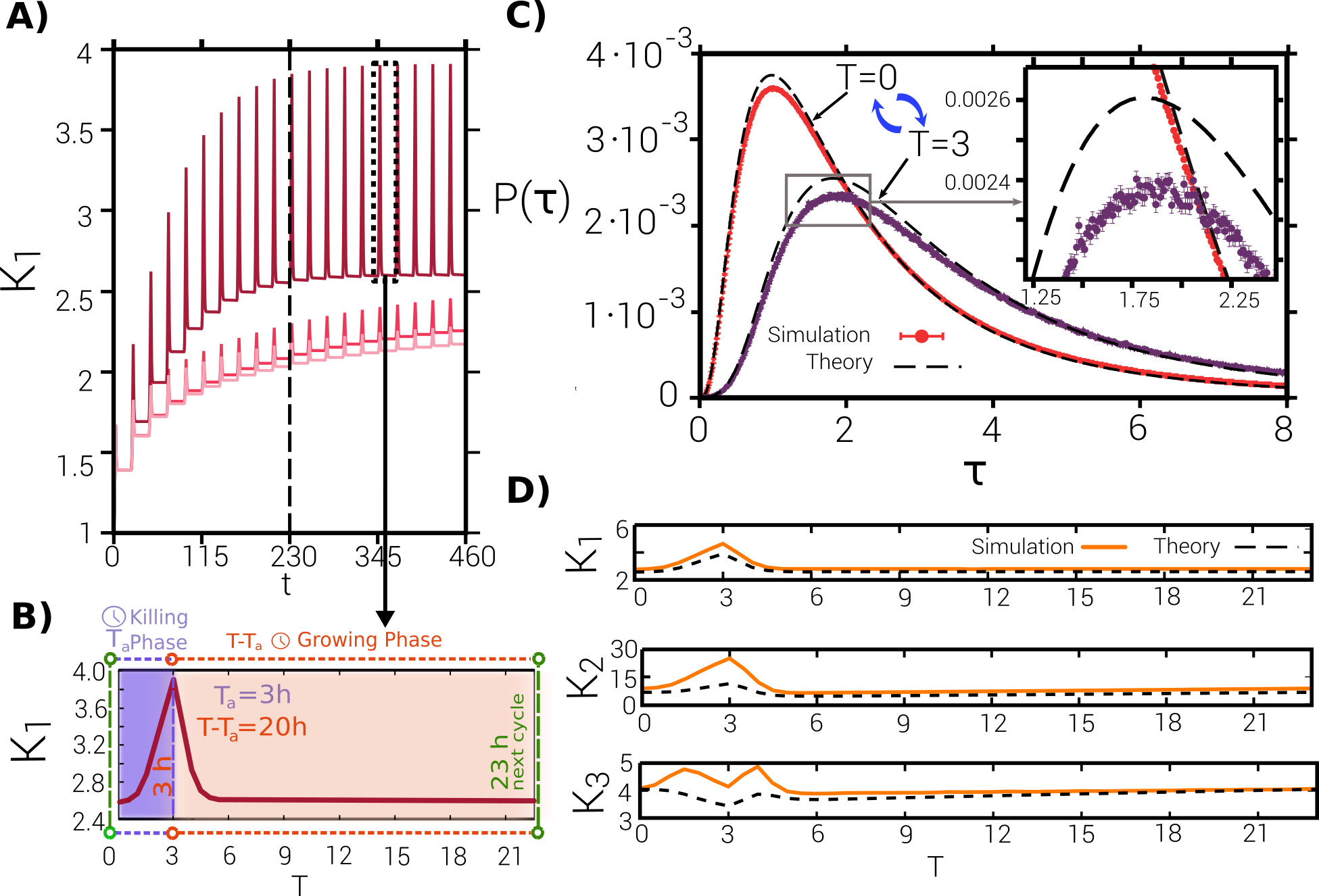}
		\caption{\small \textbf{Characterization of the asymptotic state in the
				multiplicative version of the model }.  \textbf{(A)} Approach to
			the dynamic asymptotic state for the multiplicative case, as
			resulting from the integration of Eq.(1) and Eq.(2) for $T_a= 3h$
			($t$ indicates overall time as measured in hours). The different
			curves correspond to three different values of the variation
			amplitude (\newr{from bottom to top:}
			$\alpha_{M}=0.0048, 0.01, 0.048$). The difference between this
			plot and Fig 2 is that $K_1$ is measured at different times
			within the cycle and not just right at the end of antibiotic
			exposure (see {(B)}). The vertical dashed line marks the $10th$
			cycle at which the experiment stopped.  Observe that the
			steady-state mean value, the oscillations amplitude, and the
			relaxation time depend on the variation amplitude
			$\alpha_M$. \textbf{(B)} Mean lag time within a single cycle
			($T \in [0,23] h.$) in a asymptotic state. During the killing
			phase (antibiotic exposure), i.e. for $t<T_a$, the mean lag time
			increases to maximize the number of dormant individuals; then, in
			the fresh medium the mean relaxes back to the initial
			value. \textbf{(C)} Lag-time probability distribution ---as
			derived from theory (dashed lines) and computationally (solid
			lines)--- at the start of the cycle (leftmost curve) and when
			antibiotics are removed (rightmost curve); in the asymptotic state
			the system oscillates between these two limiting probability
			distributions, both of them exhibiting heavy tails. \textbf{(D)}
			Evolution of the three first cumulants $K_1$, $K_2$, $K_3$ (mean,
			variance, and skewness, respectively) within a asymptotic cycle
			(both theoretical and computational results are shown).  Observe
			in (C) and (D) that the theory correctly predicts the properties
			of the distribution but there are some small errors due to finite
			size effects. } \label{Fig 4} \end{figure}
	
	\begin{figure}
				\centering \includegraphics[width=\textwidth]{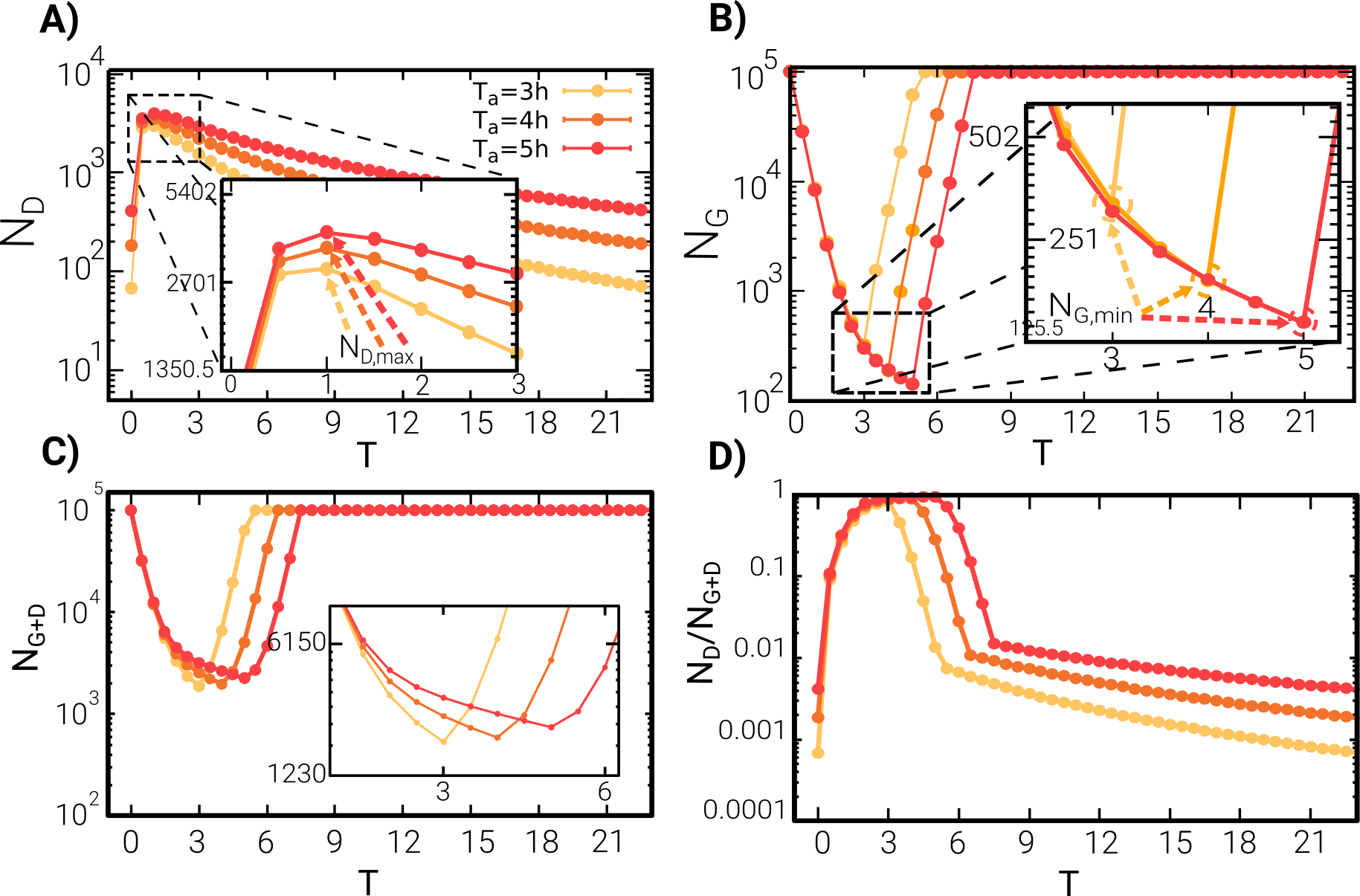} 
		\caption{\small \textbf{Population dynamics }. \newr{ Abundances $N_D$ and
				$N_G$ of dormant \textbf{(A)} and growing \textbf{(B)} cells,
				respectively, along a full cycle in the asymptotic regime (reached
				after only three cycles) for the
				multiplicative model (the curves are the result of averaging over
				many independent realizations for different $T_a$ as color coded;
				observe the semi-logarithmic scale). Dormant cells abundances
				reach a maximum value after approximately one hour of exposition
				to antibiotics, almost independently of $T_a$, and then start a
				slow decrease, while $N_G$ exhibits an opposite
				trend: it rapidly decreases and reaches a minimum at $T=T_a$
				(see inset), after which it grows exponentially fast until the
				carrying capacity is reached. \textbf{(C)} The total number of
				cells $N=N_G+N_D$ is plotted along the cycle: for all values of
				$T_a$ the absolute minimum is reached near $T_a$ (as clearly
				seen in the inset). \textbf{(D)} The fraction of dormant cells
				relative to the total number is maximal nearby $T_a$ and
				decreases when antibiotics are removed.}
		}  \label{Fig 5} \end{figure}
	

	\subsection*{Asymptotic state}
	Even if experimental results are available for a fixed and limited
	number  ($10$) of antibiotic-exposition cycles, the already-calibrated
	model allows us to scrutinize the possible emergence of asymptotic
	states after a much-larger number of cycles.  In other words, it is
	possible to go beyond the experimental limits and analyze the fate of
	the population. In this sense, the experimental results can be seen as
	a ``transient adaptation'' to the environment, while the evolutionary
	cycle would be completed only when an asymptotic (evolutionary stable)
	state is reached.  Let us remark that the asymptotic state is
	necessarily a \emph{periodic} one, as the phenotypic distributions
	vary at different instants of the cycle, i.e. the asymptotic
	distributions ---measured at arbitrary times within the cycle---
	exhibit periodic oscillations in its shape, tracking the perpetual
	environmental cyclic changes.  This is illustrated in Fig
	\ref{Fig 4}, showing results obtained by numerically
	integrating the macroscopic equations, Eq.(1) and Eq.(2). First of all,
	it shows periodic oscillations of the mean lag time $K_1$; as shown in
	panel (A) it first increases from its initial value $K_1=1$ and then,
	eventually, reaches an oscillatory steady state.  More specifically,
	as clearly seen in the zoomed plot of panel (B), within the steady
	state, the maximum mean value within each cycle is reached right
	before antibiotics removal.  This is an expected result as in the
	first part of the cycle, i.e. during the ``killing phase'', the
	presence of antibiotics induces a selective pressure towards
	increasing the mean lag-time value because delaying the exit from the
	lag phase provides protection from the antibiotics. On the other hand,
	in the fresh medium (growing phase) the selective pressure quickly
	reduces the mean lag time to foster fast growth and increased fitness.
	Thus, summing up, the periodic alternation of environmental conditions
	induces a stable periodic change in the mean lag-time value.
	
	Actually, it is not only the mean that changes periodically, but the
	whole probability distribution that varies cyclically. This is
	illustrated in Fig \ref{Fig 4}C and \ref{Fig 4}D which shows computational
	and theoretical results for the lag-time probability distribution and
	its first cumulants, $K_1, K_2$ and $K_3$, for the multiplicative case
	(similar plots for the additive case are shown in S1 Text, Sec. S6).  Observe,
	in particular, in panel C, that the distribution oscillates between
	two extreme or limiting cases corresponding to the times of
	antibiotics inoculation and antibiotics removal, respectively. This
	effect can be more vividly seen in the \new{S1 and S2 Videos}.
	
	Let us also highlight that the probability distributions exhibit
	non-Gaussian tails and are right-skewed. In particular, to make these
	observations more quantitative, Fig \ref{Fig 4}D shows the
	variance, $K_2$, and the skewness, $K_3$, along the cycle in the
	steady state. Notice also the very-good ---though not perfect---
	agreement between computational results and theoretical estimates
	(dashed lines in Fig \ref{Fig 4}C and \ref{Fig 4}D).
	
	Furthermore, let us emphasize that, importantly, the amplitude of
	the variations ---as controlled by the parameter $\alpha_M$ (or,
	similarly, $\alpha_A$ for the additive case)--- has a non-trivial
	effect on both the transient and the asymptotic behavior.  In
	particular, the value of such amplitude not only affects the mean
	value of lag times after $10$ cycles ---as illustrated by the plateau
	of the oscillations in Fig \ref{Fig 4}A and \ref{Fig 4}B--- but also (i)
	its asymptotic value, i.e.  the mean lag time, (ii) the amplitude of
	the oscillations across a cycle in the steady state, and (iii) the
	relaxation time to the asymptotic state (i.e., the speed of
	evolution).  This is due to the heavy tails of the distribution:
	increasing the amplitude of variations directly increments the
	variance of lag times, but this also enlarges the left-skewness of the
	distribution, feeding-back to the mean value. Therefore, the
	eco-evolutionary attractor is shaped both by selection and mutation,
	departing from the classical evolutionary scenario, as e.g. in
	adaptive-dynamics, in which the amplitude of the variations just
	affects the variance of the resulting distribution but not the overall
	attractor.

	Finally, we complement our observations with the population structure
	dynamics, i.e. proportion, minimum and maximum of dormant and awake
	cell numbers, in Fig \ref{Fig 5}. In particular, panels (A)
	and (B) show the abundances of dormant and awake cells as function of
	time along an asymptotic cycle. Observe that the number of dormant
	cells reaches a maximum, $N_{D,max}$, after one hour, independently of
	the antibiotic duration time $T_a$, while its height is proportional
	to this parameter. On the other hand, the position of minimum of
	growing cells number, $N_{G,min}$, scales with $T_a$ and its magnitude decreases correspondently. In Fig \ref{Fig 5} we also show and
	discuss the dependence the total number of cells $N=N_G+N_D$ (panel C)
	as well
	as the relative fraction of dormant individuals along a full cycle in
	the stationary state (which is reached after onley a few (three)
	antibiottic cycles).

	For the sake of completeness, let us also emphasize that both
	versions of the model are able to generate $MDK_{99}$ values that
	grow as a function of the number of antibotic cycles, converging to
	an asymptotic-state value; at the end of the tenth cycle simulations
	compare well with empirical observations for different values of
	$T_a$ (see \new{S10 and S11 Figs}; observe that the largest difference
	appears for $T_a=8$, a case for which also $K_1$ deviates slightly
	from $T_a$ in the experiments).

	\subsection*{Deviations between theory and simulations: finite-size effects}
	
	Thus far, we have reported results stemming from computational
	analyses of the individual based model as well as from numerical
	integration of the associated macroscopic theory, i.e. the GCK
	equation.  Small but systematic discrepancies between theory and
	simulations are evident, see for example Fig \ref{Fig 4}C and \ref{Fig 4}D.
	Let us here discuss the origin of such differences.
	
	The theoretical approach relies on two different approximations: (i)
	on the one hand it considers the small-variation approximation to
	include just the first two moments of the variation function (i.e. a
	diffusion approximation); (ii) on the other hand, in order to derive
	the macroscopic GCK equation, one needs to neglect correlations
	between individuals, a type of mean-field approximation that, as
	usual, is expected to be exact only in the \emph{infinite
		population-size limit} \cite{vKampen,Gardiner}. In S1 Text, Sec. S4, we
	show computational evidence that the small mutation approximation is
	not a significant source of errors; hence, the discrepancies
	necessarily stem from finite-size effects.  Indeed, in the present
	experimental set up, there is a bottleneck at the end of each
	antibiotics cycle, when there is a small number of surviving
	individuals, thus limiting the validity of the mean-field
	approximation in such a regime. As a matter of fact, one can clearly
	see from Fig \ref{Fig 4}D that the largest discrepancies
	appear around the end of the killing phase, when the population is the
	smallest.  Note also that the main features of the dynamics in
	phenotypic space are reproduction and variation: i.e., offspring are
	similar to their progeny. But reproduction events occur only within
	the awake (growing) sub-population; the full population-size,
	involving also dormant ones, is not the most relevant quantity to
	gauge finite-size effects. Therefore, in order to minimize the
	discrepancies between theory and simulations it does not suffice to
	consider larger population sizes: even for huge values of the carrying
	capacity $K$, we find that the population at the end of the killing
	phase is always rather small and, hence, exposed to large demographic
	fluctuations, i.e. to finite-size effects.
	
\begin{figure}
				\includegraphics[width=\textwidth]{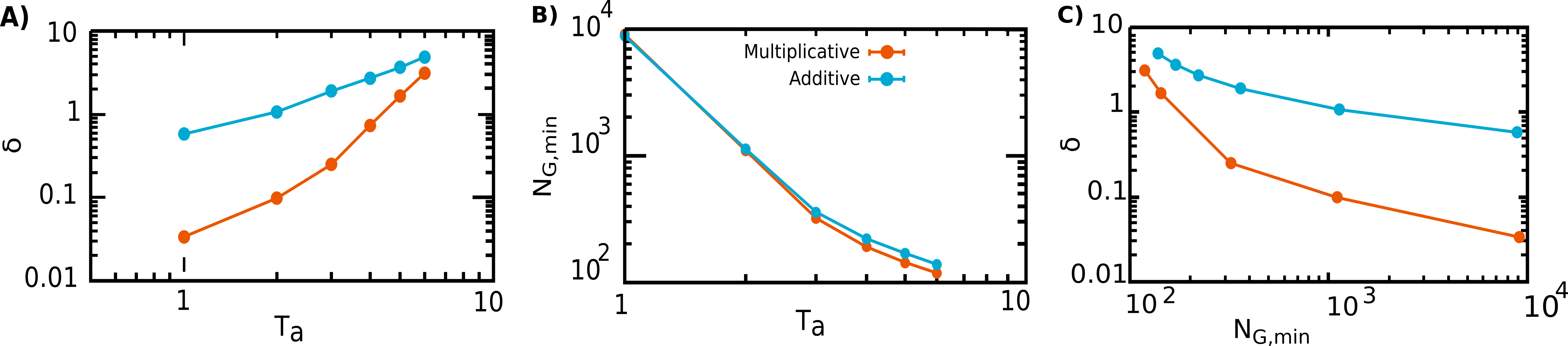}
		\caption{\textbf{Analysis of the deviations between simulations and
				theory.} The parameter $\delta$ is defined as the difference
			between the mean lag times (right at the end of the antibiotic
			cycle) in the theoretical approach and in computer simulations.
			\textbf{(A)} Double-logarithmic plot reporting the dependence of
			the error parameter $\delta$ on the antibiotic time exposure $T_a$
			for both the additive (blue dots) and the multiplicative (red
			dots) versions of the model; in either case, the larger the
			exposure time the larger the error.  \textbf{(B)} Minimum number
			of awake individual during the cycle in the asymptotic regime,
			$N_{G,min}$, as function of $T_a$ in double-logarithmic scale. As
			expected, the larger the exposure time the smaller the number of
			surviving individuals.  \textbf{(C)} Combining the data from (A)
			and (B) it follows that $\delta$ decreases with increasing
			$N_{G,min}$, meaning that deviations between theory (expected to
			be exact for infinitely large population sizes) and computational
			results stem from finite-population-size effects. Notice that
			errors are smaller in the multiplicative version of the
			model.}\label{Fig 6}
	\end{figure}

	To put these observations on more quantitative bases, we define a
	parameter $\delta$ as the deviation between theoretical and
	computational results for the mean lag-time value after antibiotic
	exposure and monitor its dependence on the minimal size of the awake
	population (i.e. right at the end of the antibiotic phase).  Fig
	\ref{Fig 6} illustrates that: (A) the deviation grows with the
	antibiotic-exposure time $T_a$, whereas (B) the minimum awake
	subpopulation size decreases with $T_a$. Combining these two pieces of
	information one can see (C) that the deviation parameter $\delta$
	decreases as the minimum subpopulation of awake individuals increases.
	Unfortunately, the convergence to zero of this last curve is very
	slow, and thus, it is computationally very expensive to remove
	finite-size effects.
	
	Finally, let us remark that we leave for future work the formulation
	of an extension of the mathematical theory accounting for finite-size
	effects \cite{Frey2011,Frey2012,Wakano}, including corrections to the
	GCK equation.

	\section*{Conclusions, discussion and perspectives}
	
	\vspace{1cm}
	
	{\bf Summary of results and conclusions.} We have presented a
	mathematical and computational model to quantitatively analyze the
	emergence and evolution of tolerance by lag in bacteria. Our first
	goal was to reproduce the main results reported in the laboratory
	experiments of Fridman \emph{ et al.} \cite{Fridman} in which the
	authors found a very fast evolution of tolerance by lag in a community
	of \emph{Escherichia coli} bacteria periodically exposed to an
	antibiotics/fresh-medium cycle.  In particular, after a relatively
	small number of such cycles, \newr{there is a clear change in the individual-cell lag-time
		distribution with its  mean value evolving to
		match the duration of antibiotic exposure.}  This is remarkable, and
	demonstrates that tolerance by lag is the first and generic strategy
	adopted by bacteria to survive under harsh environmental conditions
	such as the presence of antibiotics. A second key empirical finding is that
	concomitantly with the evolution of the mean lag time, also the
	variance of lag times \newr{is significantly increased for longer
		antibiotic-exposure periods: i.e. the harsher the conditions the more
		diversified the lag times within the population.}  More generally, the full
	lag-time distribution becomes wider and develops a heavy tail for
	\newr{sufficiently large times}.  This means that there exist
	individual phenotypes that are clearly sub-optimal under the strictly
	controlled laboratory conditions and most-likely reflects a
	bet-hedging strategy, preparing the community to survive under even
	harsher conditions (i.e. longer stressful periods).
	
	To shed light onto these observations we developed a stochastic
	individual-based model assuming that individuals are characterized by
	an intrinsic lag time, setting the ``typical'' time at which such
	individual stochastically wakes up after dormancy. This phenotypic
	trait is transmitted to the progeny with possible variation. By
	considering a protocol analogous to the experimental one
	(i.e. alternating antibiotic exposure and fresh medium growth) the
	model is able to produce a distribution of characteristic lag times
	across the population that reproduces quite well the empirical results
	in all cases by tuning a single parameter value.  In particular, the
	emerging lag-time distributions have a mean that matches the period of
	antibiotic exposure $T_a$, an increase of the mean and variance with
	$T_a$, as well as a large difference between the mean and the median,
	which result from the appearance of heavy tails in the lag-time
	distributions. Nevertheless, it is important to underline
	that the distributions that the model generates are just a proxy for
	the empirically-determined ones, where the actual times in which
	individual bacteria give rise to new and detectable (i.e. visible
	with the available technology) colonies are measured.
	
	Importantly, in order to account for all the above \newr{empirical
		phenomenology}, the model needs to assume \emph{multiplicative
		variations}, i.e.  that the variability between the parent's trait
	and those of its offspring increases (linearly) with the parent's lag
	time: the larger the parent's lag time the larger the possible
	variation. This multiplicative process --- at the roots of the
	emerging heavy tails in the lag-time distribution--- resembles the
	so-called \emph{rich-get-richer} mechanism of the \emph{Matthew
		effect} \cite{Sornette, Mitzenmacher, Newman, MN-Munoz,
		Barabasi}. This type of variations implements an effective dependence
	between the parent's trait value and the variation amplitude, that was
	hypothesized as a possible mechanism behind the experimental results
	and that could stem from a highly non-linear map between genotypic
	changes and their phenotypic manifestations
	\cite{Fridman,Toxin_Balaban}.
	
	Notably, our analyses reveal that the amplitude of variations affects
	not just the variance ($K_2$) of the resulting lag-time distribution,
	but also its mean ($K_1$) as well as other higher-order cumulants such
	as the skewness ($K_3$). This is in blatant contrast with standard
	approaches to evolutionary or adaptive dynamics, in which the
	``mutational amplitude'' only influences the ``broadness'' ($K_2$) of
	the distribution of traits in phenotypic space, but does not alter the
	attractor of the dynamics (e.g. $K_1$). Thus, the introduction of
	state-dependent (multiplicative) variability constitutes a step
	forward into our understanding of how simple adaptive/evolutionary
	processes can generate complex outcomes.
	
	Let us finally mention that our model describes rapid evolution, where
	ecological and mutational time scales are comparable.  This interplay
	between ecological and evolutionary processes is explicit in the
	asymptotic state: it is not an ``evolutionary stable state'' but a
	\emph{``non-equilibrium evolutionary stable state.'' }  By
	non-equilibrium we mean that the detailed-balance condition ---a
	requirement of equilibrium states \cite{Gnesotto_2018}--- is violated
	and thus, there are net probability fluxes in phenotypic space. These
	correspond to adaptive oscillations in phenotypic space. Key
	properties of such a state (oscillation plateaus, amplitudes, etc.)
	depend on the mutational amplitude, i.e., the amplitude of variations
	determine the eco-evolutionary attractor.  In future work we will
	scrutinize much in depth non-equilibrium characteristic properties,
	such as non-vanishing entropy-production of these type of complex
	eco-evolutionary processes \cite{Mustonen4248, Leibler-Kussell,
		Kussell2014}.
	
	\vspace{1cm} {\bf Advantages and limitations of the
		phenotypic-modeling approach.}  As already underlined, the present
	model assumes adaptation at a phenotypic level.  Is this a
	biologically realistic assumption? The answer to this question, in
	principle, is affirmative but some caveats are in order.
	
	First of all, let us recall that a large part of the theoretical work
	on evolutionary dynamics and adaptation developed during the last
	decades focuses on phenotypic adaptation. For instance, in the theory
	of adaptive dynamics, individuals are always characterized by some
	phenotypic trait or set of traits which is subject to selection and
	transmission to the progeny with variation \cite{Geritz1,Geritz2,
		beginners,DD1,DD2} (see also e.g. \cite{Bonachela1,Bonachela2}). In
	general, this is the most parsimonious way of modeling adaptation as
	the details of the genotypic-phenotypic mapping are usually highly
	non-linear or simply unknown (see
	e.g. \cite{Pascual,Dalziel,Manrubia2014,Manrubia2017,Manrubia2020}).
	
	On the one hand, adaptation beyond genetic changes ---for example
	epigenetic adaptation--- is a well-documented phenomenon in the
	bacterial world \cite{Novick1957} and is the focus of intense research
	activity
	\cite{Casadesus1,Casadesus2,Kuipers,Balaban-epigenetic,phenotypic-memory}. For
	instance, recent work explores ``the evolutionary advantage of
	heritable phenotypic heterogeneity'', which suggests that evolutionary
	mechanisms at a phenotypic level, such as the ones employed in our
	approach, might be biologically favored with respect to more-standard
	genetic mechanisms, under certain circumstances \cite{Carja2017}. In
	particular, such \newr{phenotypic variability} can provide a faster and more flexible type of
	response than the one associated with traditional genetic mutations.
	
	Nevertheless, it is important to underline that Fridman \emph{et al.}
	found empirical evidence that ---in their specific setup--- genetic
	mutations were always present in the evolved strains. In particular,
	they found mutations in genes controlling the so-called
	toxin-antitoxin circuit, mediating the response to antibiotic stress
	\cite{Fridman}. This regulatory circuit is known to lead to
	``multiplicative fluctuations'' in the lag-time distribution at the
	phenotypic level \cite{Toxin_Balaban}. Thus, strictly speaking, our
	modeling approach constitutes an effective or phenomenological
	approximation to the more complex biology of this problem.
	
	\newr{This observation opens promising and exciting avenues for future
		research} to shed light on how broad probability distributions of lag
	times ---possibly with heavy tails--- can be actually encoded in
	phenotypic or genetic models. Actually, scale-free (power-law)
	distributions of bacterial lag times have been recently reported in a
	specifically-devised experimental setup \cite{powerlaw}. Similarly to
	our conclusions, this work also emphasizes that a broad distribution
	of individual-cell waking-up rates is needed to generate
	non-exponential decays of the overall lag-time distribution.
	
	Similarly, another exciting possibility would be to develop
	computational models akin to the phenotypic one proposed here but
	implementing genetic circuitry; i.e. models where the phenotype is the
	(possibly stochastic) outcome of an underlying regulated genetic
	process and where the object of selection are not specific lag times
	but their whole distributions as genetically encoded.

	
	\vspace{1cm}
	
	{\bf Future developments and perspectives.} In future research, we
	would like to further delve onto several aspects, both biological and
	theoretical, of the present work. As a first step, we leave for
	forthcoming work the analysis of the pertinent question of how similar
	systems respond to randomly fluctuating environments as opposed to
	periodically changing ones; do they develop heavier tails to cope with
	such unpredictable conditions in a sort of bet-hedging strategy?
	How do the statistical features of the environmental variability translate
	into the emerging lag-time distributions?
	\cite{Kussell-Leibler,Kuipers,Mitarai,hidalgo,Villa2,Villa1}.
	
	From a more theoretical perspective, we leave for an impending work the
	formulation of an extension of our approach that fully accounts for
	finite-size effects, thus introducing the next-to-leading order
	corrections to the generalized-Crow-Kimura macroscopic equation
	accounting for demographic fluctuations.  Within this context, treating the
	variation-amplitude itself as an evolving trait is also a potentially
	fruitful route for further studies.
	
	Finally, as a long-term project we plan to develop models and
	analytical approaches, similar to the ones explored here, but focusing
	on genetic evolution, employing explicit genotypic-phenotypic
	mappings, rather than just on phenotypic changes. In particular,
	by introducing this further layer of complexity it would be possible
	to generate more general types of single-cell lag-time distributions,
	not limited to exponential ones as the purely Markovian approach
	considered here. Let us recall that a more general stochastic
	non-Markovian framework ---i.e., including memory effects (see
	e.g. \cite{Norman-review,Mitarai,Zhang})--- is a
	challenging goal that promises to be very pertinent and relevant for
	many diverse problems in which the control of time is important.

	\begin{small}
		\section*{Methods}\label{ch:methods}

		\subsection*{Numerical values of the parameters}\label{ch:sim_detail}
		In order to fix parameter values we employed the experimental values
		and measurements in \cite{Fridman} as closely as possible. The number
		of bacteria involved in the experiment reaches values of the order of
		$\sim 10^9$; however this number is prohibitively large for computer
		simulations and we fixed a maximum carrying capacity of $K=10^5$,
		verifying that results do not depend strongly on such a choice (see
		finite-size effects section). Initially the number of cells in the
		growing state is fixed to be equal to the carrying capacity $K$; thus
		no cell is initially in the dormant state). The doubling time of both
		the ancestral and the evolved populations is $25\pm 0.3 \min$; thus on
		average every single bacteria attempts reproduction at a rate
		$b=1/25 \min= 2.4 \ h^{-1}$. The death rate for (natural) causes
		(i.e. other than antibiotics) is $d=3.6\cdot10^{-5} \ h^{-1}$. The
		awakening rate is given by the inverse of the characteristic time
		$a=1/\tau$ \cite{Balaban}. The initial condition (ancestral or wild
		population) was randomly sampled from a truncated Gaussian peaked at
		$\tau=0$. Since the empirical ancestral distribution is narrow and
		close to the origin \cite{Fridman} (mean lag time
		$\langle \tau \rangle^{exp.}_0 =\ 1\ \pm0.2\ h$) we fix the standard
		deviation of the truncated Gaussian distribution to
		$\sigma=1\ h\ 16\ \min$ in such a way that
		$\langle \tau \rangle^{sim.}_0\sim1\ h$. Neither the exit rate from
		dormancy $s$ nor the amplitude of the mutations, $\alpha_A$ and
		$\alpha_M$, can be experimentally measured, but we can fix them
		indirectly (the rest of the parameters are kept fixed with the values
		specified above). First, $s$ can be chosen using the experimental
		information that for the ancestral population $MDK_{99}\sim 2.55
		h$. Hence, we leave the (simulated) ancestral population in the
		antibiotic phase until the $99\%$ becomes extinct; averaging over
		different initial conditions, we found that $s=0.12\ h^{-1}$ is a good
		approximation.  To fix the constants $\alpha_A$ and $\alpha_M$ we
		performed simulations for diverse values of such parameters and looked
		for those that best reproduce the experimental tendency after $10$
		exposure cycles for the different exposure times under consideration
		(in particular, we performed a least-square deviation analysis to
		match the straight-line $\langle \tau \rangle=T_a$ when performing a
		linear interpolation for all $T_a$'s). A systematic sweep of the
		values of the first two significant digits led us to $\alpha_A=0.16 h$
		and $\alpha_M=0.048$.
		
		\subsection*{Variation Functions}
		We consider two different variation kernels for lag-time variations
		$\delta=\tau- \tilde{\tau}$: the additive one,
		$\beta_{A}(\delta;\tilde{\tau})$ and multiplicative one,
		$\beta_{M}(\delta;\tilde{\tau})$. Both of them are probability density
		functions of $\delta$, normalized in $[-\tilde{\tau},\infty]$ and may
		depend on the initial phenotype $\tilde{\tau}$.  In particular, the
		additive case reads:
		$
		\beta_{A}(\delta;\tilde{\tau})=e^{-\frac{\delta^2}{2\alpha_{A}^2}}/Z_{A}(\tilde{\tau})$
		with
		$Z_{A}(\tau)=\alpha_{A}\frac{\sqrt{\pi}}{\sqrt{2}}Erfc(-\frac{\tau}{\sqrt{2}\alpha_{A}})$
		(where $Erfc$ stands for the complementary error function), while in
		the multiplicative case, we consider
		$\beta_{M}(\delta;\tilde{\tau})=e^{-\frac{\delta^2}{2\alpha_{M}\tilde{\tau}}}/Z_{M}(\tilde{\tau})$
		with
		$Z_{M}(\tau)=\alpha_{M}\tau\sqrt{\frac{\pi}{2}}Erfc(-\frac{1}{\sqrt{2}\alpha_{M}})$.
		
		\subsection*{Measuring lag-time distributions}
		In order to determine lag-time distributions, we computed histograms
		in phenotypic space, as discretized in bins of size
		$\Delta\tau = 10^{-2}$ and averaged over many realizations of the
		process.  In the asymptotic steady state, similar histograms were
		computed at different times along the antibiotic/fresh-medium cycle
		(e.g. right after antibiotic inoculation or after antibiotic
		removal). To obtain results for the transient state we determined the
		histogram after running for $10$ cycles. On the other hand, to
		determine the steady state, we started measuring after $300$ cycles
		(to make sure that a steady state has been reached) and then collect
		statistics up to cycle $1500$, at intervals of $10$ cycles to avoid
		correlations. We repeated the process for $30$ realizations and
		calculated the histogram as well as the associated cumulants.
		
		\subsection*{Numerical Integration of the macroscopic equation} 
		The parameter set and initial condition for numerical integration of
		the mean-field equations are the same as specified above. Numerical
		integration was carried out using the finite differences method. In
		particular $\partial_t$ was approximated using first order forward
		differences, $\partial^2_\tau$ using second-order centered
		differences, and integrals were approximated as Riemann sums
		\cite{Pearson}. The numerical integration steps used in the figures
		are: $h_t=10^{-6}$ and $h_\tau=10^{-2}$ . Note that, when we calculate
		the probability distributions during the simulation, we must use the
		same bin size to be able to correctly compare with the theoretical
		distributions later. We used absorbing boundary conditions
		$\phi(\tau=0,t)=\phi(\tau=\tau_{max},t)=0$, where $\tau_{max}$ is the
		limit of the phenotype space considered for the numerical integration,
		in particular: $\tau_{max}=60h$.
		
		\vspace{2cm} {\bf Acknowledgments:} We are grateful to N. Balaban for
		sharing data with us as well as for her very valuable insights and
		comments. We also thank R. Rubio de Las Casas, S. Manrubia,
		G. B. Morales, J.A. Bonachela, J. Grilli, and V. Buendía for
		stimulating and fruitful discussions.

	\end{small}

	\vspace{1cm}
	{\bf  SUPPORTING INFORMATION:}
	\vspace{1cm}
	
	\new{\textbf{S1 Text}}: Sections: S1: The Microscopic process, S1A: Master equation, S1B: Gillespie algorithm. S2: From Microscopic to Macroscopic Process: S2A. Marginalization, S2B: Mean Field Approximation. S3: Small Variation Approximation. S4: Deviation between theory and simulations, S4A: Validity of the small variation approximation and border effects, S4B: FInite-size effects. S5: Spontaneous shifting to the dormant state. S6: Additional Figures. S7: Movies. 
	
	\vspace{0.5cm}  \new{\textbf{S1 Fig:}} \textbf{Validity of the small variation approximation in the additive case.} Evolution of the first three cumulants ($K_1,K_2,K_3$) in a asymptotic cycle calculated  by the numerical integration of  both the general mean-field equation (S95)(solid line)
	and the generalized Crow-Kimura equation (S114)(points) for two different values of $\alpha_A$,  in \textbf{(A)} $\alpha_A=0.16h$ and in \textbf{(B)} $\alpha_A=0.035h$. The first variation value is the  best fit the experimental results (main text), while  the second causes negligible border effects. In both cases the small-variation approximation is a good approximation. Parameter values: $T_a=3h$, the rest of the parameters, as well as the initial conditions, are kept fixed as specified in the main text).
	
	\vspace{0.5cm}  \new{\textbf{S2 Fig:}} \textbf{Range of validity of the small variation approximation in the additive scenario}.
	Systematic comparison of Eq.(S114) and Eq.(S95) via parameter  Eq.(S124) for different $\alpha_A$ values, both axis in $log-$scale. Note that the deviation monotonically increases with $\alpha_{A}$. In particular  for $\alpha_{A}=0.16 h$, the one used in the main text results, the deviation $\delta_{st.}=(3.5 \pm 0.2)\cdot 10^{-2} h$, Eq.(S124)  is small enough to use Eq.(S114). Parameter values: $T_a=3h$, the other as in the main text.
	
	\vspace{0.5cm}  \new{\textbf{S3 Fig:}} \textbf{Validity of the small variation approximation in the multiplicative scenario}.  Evolution of the first three cumulants ($K_1,K_2,K_3$) in a asymptotic cycle calculated  by the numerical integration of  both the general mean-field Eq.(S95)(solid line)
	and the generalized Crow-Kimura Eq.(S114)(points). The generalized Crow-Kimura equation is a good approximation of the general mean-field equation for the $\alpha_M$-value that best fits the experimental results, e.g.$\delta_{st.}=(1.5 \pm 0.3)\cdot 10^{-3} h$.  Parameter values: $T_a=3h,\alpha_M=0.048$, the remaining are specified in the main text.
	
	\vspace{0.5cm}  \new{ \textbf{S4 Fig:}} \textbf{Border effects in the variation functions}. First (\textbf{A}-\textbf{B}) and second (\textbf{C}-\textbf{D}) moment of the variation functions as function of the trait $\tau$ in additive (left) and multiplicative (right) variation cases. In \textbf{(A)}   the additive case,   $\theta_A (\tau)$ is positive in $\tau=0$ but decays rapidly to zero as  $\tau$ increases, such that it is sufficient to restrict $\tau$ axis between 0 and 1. In \textbf{(C)} the same is shown for the second moment $\sigma_A^2(\tau)$. Note that the  magnitude of the dependence  decreases with $\alpha_{A}$).
	Nevertheless, the value used in the main text, $\alpha_A=0.16h$,   is too large to neglect Eq.(S120) this effect in the generalized Crow-Kimura eq. On the other hand, in the multiplicative case   the main text value, $\alpha_M=0.048$, is  small enough to avoid the border effects in the GCK eq.  In \textbf{(B)} one can observe that  $\theta_A (\tau)$ almost vanishes, and in \textbf{(C)} that the exact moment $\sigma^2_{M}(\tau)$ coincides quite well with the approximation  $\alpha_M\tau^2$, Eq.(S123).
	
	\vspace{0.5cm}  \new{\textbf{S5 Fig:}} \textbf{Schematic definition of the Parameter $\delta$} (illustrated in the multiplicative amplitude scenario). $\alpha_M=0.048, T_a= 6h$
	
	\vspace{0.5cm}  \new{\textbf{S6 Fig:}} \textbf{A)} Functional dependence of the rate to enter the dormant state in fresh medium $s_f$ and the number of particles in growing state $N_G$. \textbf{B)} and \textbf{C)} Lag-time probability distribution function, $P(\tau)$, at the end of the tenth cycle (for the multiplicative case)  for a constant $s_f$ and  $s_f=s_{k}\left(\text{tanh}[-cN_G/K]\right)$, respectively. The results in the main text, for $s_f=0$, are represented by a dashed line. Observe that in both cases the value of $K_1,K_2,K_3$ increase, but the qualitative form of the distributions remains unchanged. Parameter values: $h_\tau=0.01$, $\alpha_M=0.048$, $T_a=3h$, $c=3$, $T=23h$, the rest of parameters are fixed as in the main text.
	
	\vspace{0.5cm}  \new{\textbf{S7 Fig:}} \textbf{Dynamics of the averaged population structure within one cycle}. Abundances $N_G$ and $N_D$ along the $10$th cycle for $s_f$ constant ---\textbf{A)} and \textbf{C)}--- and $s_f=\tilde{s}\left(\text{tanh}[-cN_G/K]\right)$ ---\textbf{B)} and \textbf{D)}---. The two scenarios are very similar to each other. At the beginning of the cycle $N_D$ is non-vanishing, since bacteria entered this state during the fresh phase of the previous cycle. During the antibiotic exposure phase both, $N_D$ and $N_G$, decrease to a minimum as the bacteria die. When the antibiotic is removed and a fresh medium is added, $N_G$ grows towards the system's carrying capacity. Observe that the bacteria still enter the dormant state, but the reproduction rate is much higher, in such a way that an overall reduction of $N_G$ is only observed as the system approaches the carrying capacity.
	
	\vspace{0.5cm} \new{\textbf{S8 Fig:}}
	\textbf{Characterization of the asymptotic  state in the additive version of the model.} \textbf{(A)} Relaxation of the mean lag-time to its asymptotic state (curves obtained from  the integration of Equation Eq.(S114)  with $T_a=3h$ and additive model. The different curves correspond to three different values of the variation amplitude, $\alpha_A$; from the lowest to the highest: $\alpha_A=0.035h$, $\alpha_A=0.1h$ and $\alpha_A=0.16h$.  \textbf{(B)}  Zoom of the curve $\alpha_A=0.16h$ for one single cycle. In particular, the mean lag time, $K_1$, is shown along a cycle in the asymptotic regime. At $T=0$ the antibiotic is added and the system enters in the ``killing phase'' ($T\in [0,T_a]$). When the antibiotic is present, the system experiences a selection pressure towards longer lag times, in consequence, $K_1$ increases. At $T=T_a$ the antibiotic is washed and the fresh medium is added ($T \in [T_a,T_{max}]$). In this regime, the selection pressure is towards shorter lag times and $K_1$ relaxes back to the initial value. \textbf{(C)}  Lag-time probability distribution at $T=0$ (leftmost curve) and $T=T_a=3h$ (rightmost curve) as derived theoretically (Eq.(S114), dashed lines) and computationally (dots). In the asymptotic state the system oscillates between these two limiting probability distributions, both of them exhibiting weak  tails.  \textbf{(D)} Evolution of the first three cumulants, $K_1,K_2$ and $K_3$, (mean, variance , and skewness respectively) along a cycle in the asymptotic state (both theoretical and computational results are displayed). Observe that in C/D the theory correctly predicts the properties of the distribution, however there are deviations due to finite size effects.  
	
	\vspace{0.5cm}  \new{\textbf{S9 Fig:}} \textbf{Simulated number of cells via the Gillespie algorithm} ---additive-amplitude scenario, linear scale---. \textbf{a)} Number of cell in dormant state during a whole cycle in the asymptotic state for different exposition times $T_a$. \textbf{b)} Same as a) but for growing state bacteria. Obviously, the minimum number of bacteria in growing state $N_{G,min}$ is reached at $t=T_a$, since, during the antibiotic exposure phase, that number can only decrease. Parameters: $\alpha_A=0.16 h$, the rest are fixed as in the main text.
	
	\vspace{0.5cm} \new{\textbf{S10 Fig:}} \textbf{Evolution of the $MDK_{99}$ over 10 exposure cycles}. \textbf{a)} Additive case. \textbf{b)} Multiplicative case.  In our simulations, the $MDK_{99}$ is calculated analogously to the experimental procedure.  After a certain number cycles (i.e. $\#$ in the figure) of antibiotics-fresh environment, the evolved population is posed back in the antibiotic phase for a long time. The maximum number of cycles is $10$ as in the experiments. The $MDK_{99}$ is estimated by the time necessary to kill  to the  $99$\% of the population.  Both in the additive and multiplicative case, the $MDK_{99}$ increases with the $\#$ and $T_a$. Interestingly,  in the multiplicative case by increasing  $T_a$ the change in $MDK_{99}$ is bigger than the additive one. 
	
	\vspace{0.5cm} \new{\textbf{S11 Fig:}}
	\textbf{Comparison of experimental and simulated $MDK_{99}$ of the evolved population after $10$ cycles of exposure}. For both the additive and multiplicative cases, we observe that the simulated $MDK_{99}$ falls within ---or it is very close--- the experimental  values (i.e. the mean plus error) for $T_a=3h$ and $5h$, but outside for the case of $T_a=8h$. This result is to be expected given the higher noise of the experimental measurements. In particular the experimental mean is $\langle \tau \rangle^{exp.}_{T_a=8h}=10 \pm 1 h$ higher than the theoretical prediction of $8h$.
	
	\vspace{0.5cm} \new{{\bf  S1 Video:}}
	This supporting information file contains a video showing the evolution
	of the asymptotic lag-time probability distribution for the additive
	version of the model.

	\vspace{0.5cm} \new{{\bf  S2 Video:}}
	This supporting information file contains a video showing the evolution
	of the asymptotic lag-time probability distribution for the multiplicative
	version of the model.


\end{document}